\documentclass[final,5p,times,twocolumn]{elsarticle}

 \usepackage{graphics}

\usepackage{amssymb}

 \usepackage{lineno}

\journal{NIMA}

\begin{document}

\begin{frontmatter}



\title{Flavor revolution at ICECUBE horizons?}


\author{Daniele Fargion  and Paolo Paggi }

\address{Physics Department, Rome University 1,and INFN --Pl. A. Moro 2, 00185, Rome, Italy}

\begin{abstract}
Recently (May-November 2013) highest energy  neutrino  events have been presented by ICECUBE. Most ($21$) of all these ($28$) events are cascades shower whose flux exhibits a sharp  hardening respect other lower energy  atmospheric neutrino component, events suggesting an injection of extraterrestrial neutrino, mostly $\nu_{e}$, $\nu_{\tau}$, making cascades. ICECUBE claimed that a  component ($10.6^{+5.0} _{3.6}$) of these events  must be a trace of
expected downward  muons and-or atmospheric neutrinos (mostly muon track dominated): this imply that  nearly all of the few observed muon tracks  (at least $6$ of the $7$) must be themselves of  atmospheric origin: therefore  remaining $16-18$ extraterrestrial  events must be mostly of electron or of tau flavor (or rare neutral current events). The probability that this scenario  occurs is very poor, about $ 0.1-0.5\%$.  This $\nu_{\mu}$ $\bar{\nu}_{\mu}$ paucity paradox  cannot  be  solved   if  part or even all the events are made by terrestrial prompt charmed signals, because, their probability to solve the puzzle is still below  $ 1.31\%$. The paradox  might be mitigate and somehow solved if nearly \emph{all of the 28 events} are originated  by extraterrestrial sources arriving to us in de-coherent states.
At first sight a partial flavor solution  may rise if highest energy
 events at  $E_{\nu}> 60 TeV$ ( $17$ showering versus $4$ muon tracks) are mostly of extraterrestrial nature. This solution leaves nevertheless problematic the earlier $30-60$ TeV energy region, whose $8$ showers versus $3$ tracks is in more in tension with most atmospheric neutrino signals, by a sharp difference at TeV energy ruled (as shown in Deep Core) by ten over one neutrino (muon) events over showers.
This puzzling (fast) transition from atmospheric $\nu_{\mu}$ $\bar{\nu}_{\mu}$ flux at TeV to tens TeV has deep consequences: more abundant ten TeV extraterrestrial neutrino  maps may better point to astronomical clustering  or sky anisotropy; counting vertical versus horizontal (neutrino) muons  crossing the whole ICECUBE , i.e. testing zenith anisotropy at tens TeV of muons up-going, may also better disentangle and confirm their mostly extraterrestrial (or atmospheric, respectively, more vertical and isotropic versus horizontal ones) neutrino nature. Few cascades shower events in early Antares yearly might also test the $\nu$ flavor changes above $\geq 10^{12}$ eV up to a rare one at $\simeq 3\cdot 10^{13}$ eV signal. Additional EeV $\tau$ air-shower induced by UHE  $\nu_{\tau}$ within  mountains  or Earth skin  \cite{Fargion02} while skimming  \cite{Feng02}, terrestrial ground as AUGER arrays \cite{Bertou2002} might be  rare but the correlated horizontal upward PeVs $\tau$ air-shower  may soon \cite{Aita2011} shine into ASHRA crown telescopes at mountain edges by  their Cherenkov flashes.
\end{abstract}


\end{frontmatter}

\section{Introduction: the  $\nu_{\mu}$ fast flavor metamorphosis}
Cosmic ray nuclei and nucleons  scattering on top atmosphere makes (by pions, Kaons and late muons) a  final persistent neutrino raining  called  atmospheric
neutrinos ruled at high GeVs-TeVs energies by muon flavor. The  parent charged cosmic rays are widely smeared  by solar, galactic magnetic fields. In
the same way  their secondaries, the atmospheric neutrinos, are commonly spread homogeneously in the sky. Therefore diffused atmospheric neutrinos
cannot offer any  astronomy yet,  as it has been shown by four hundred thousands neutrinos in ICECUBE. As observed and expected atmospheric
neutrinos up to TeVs energies exhibit  a muon flavor dominance, that suddenly (and somehow surprisingly) it is  overthrown by recent $28$ highest \cite{Klein-2013}
energy ICECUBE events.  Indeed the  $28$ ICECUBE events show a ruling cascade showering nature ($21$ events) and a rare muon ($7$ events) track
signature. Therefore the new break is not just in the spectra hardening but mainly it is in the sudden and remarkable $\nu_{e}$
$\bar{\nu}_{e}$,$\nu_{\tau}$ $\bar{\nu}_{\tau}$ flavor " \emph{sorpasso}" or  "surpassing over" $\nu_{\mu}$ $\bar{\nu}_{\mu}$. Let us remind that
atmospheric neutrino are born mostly after nucleon scattering by pion \emph{and} muon decays. These two way for a neutrino production  (around GeVs
energy) makes both lepton flavors, mostly  muons (twice muon over electron ones). Along the vertical axis, at a few ten of GeV energy, there is also a narrow  windows for muon neutrino oscillations and its average peculiar anisotropy that makes them halves the primary ones.

\subsection{Showering cascades surpassing $\nu_{\mu}$$\bar{\nu}_{\mu}$ tracks}

\begin{figure}[hbt]
\includegraphics[scale=.155]{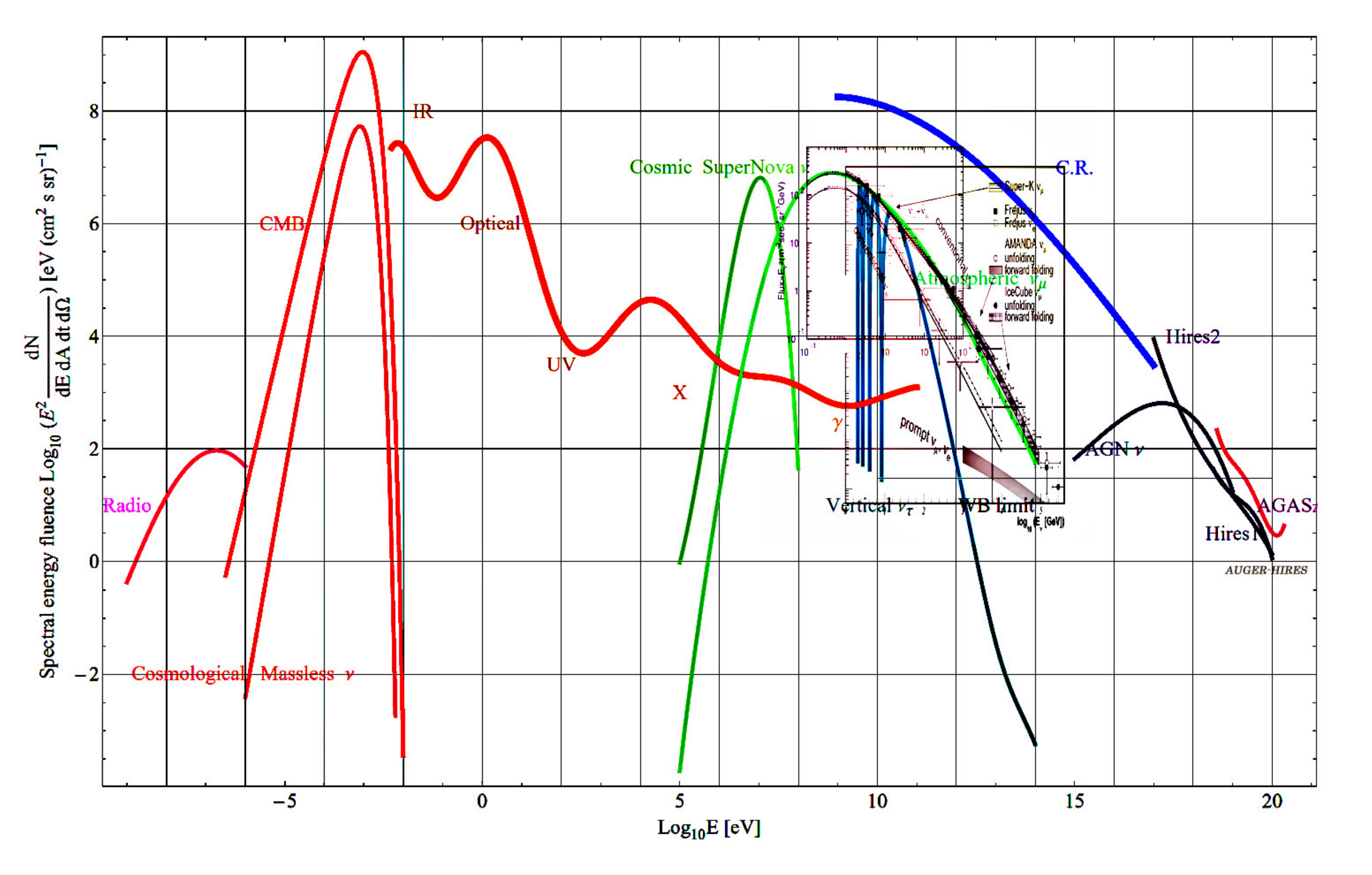}
\caption{A  fluency energy spectra overview  $\frac{dN}{dtdAd\Omega}\cdot{E^{2}}$ for all photons, cosmic rays and neutrinos in the Universe sky, in log-log scale as a function of the energy. The the atmospheric neutrino oscillation flux in the vertical axis suffer a shrinkage oscillation due to the log scale, for tau neutrino appearance; we overlap the  ICECUBE observed average atmospheric $\nu_{\mu}$ $\bar{\nu}_{\mu}$ neutrino flux, shown better in next
figures.} \label{Fig1}
\end{figure}

As soon as the muon  energy increases their relativistic  decay in flight (whose time life-distance corresponds to $0.6$ km) may overcome at few GeV
the $\simeq 12$ km atmosphere height of  vertical terrestrial  atmosphere, where shower develops. The average muon mixing suppression  is
leading to up-going events rate  half of the  primary muon  $\nu_{\mu}$ ones while the  $\nu_{e}$ $\bar{\nu}_{e}$ component rate decrease above GeVs because muons has not time to decay in flight, see Fig. \ref{Fig1}, \ref{Fig6} while pion  and kaon may still do it. The marginal vertical $\nu_{\mu}$ suppression at GeVs energies is  due to flavor mixing along the Earth that deplete  muon neutrino converting them into nearly unobservable $\nu_{\tau}$ (hardly observable  because of the large mass ${\tau}$ mass threshold). The same $\nu_{\mu}$ oscillation into  $\nu_{\tau}$ become severe at $20$ GeV at vertical axis, because of  a last complete oscillation; at higher energy $\nu_{\mu}$ $\bar{\nu}_{\mu}$ have no time of flight to mix within Earth size; therefore $\nu_{\mu}$ $\bar{\nu}_{\mu}$ keep their original flavor above hundreds GeV becoming soon twice the electron flavor and even more at higher energy because of suppression of muon decay.

\begin{figure}[hbt]
\includegraphics[scale=.41]{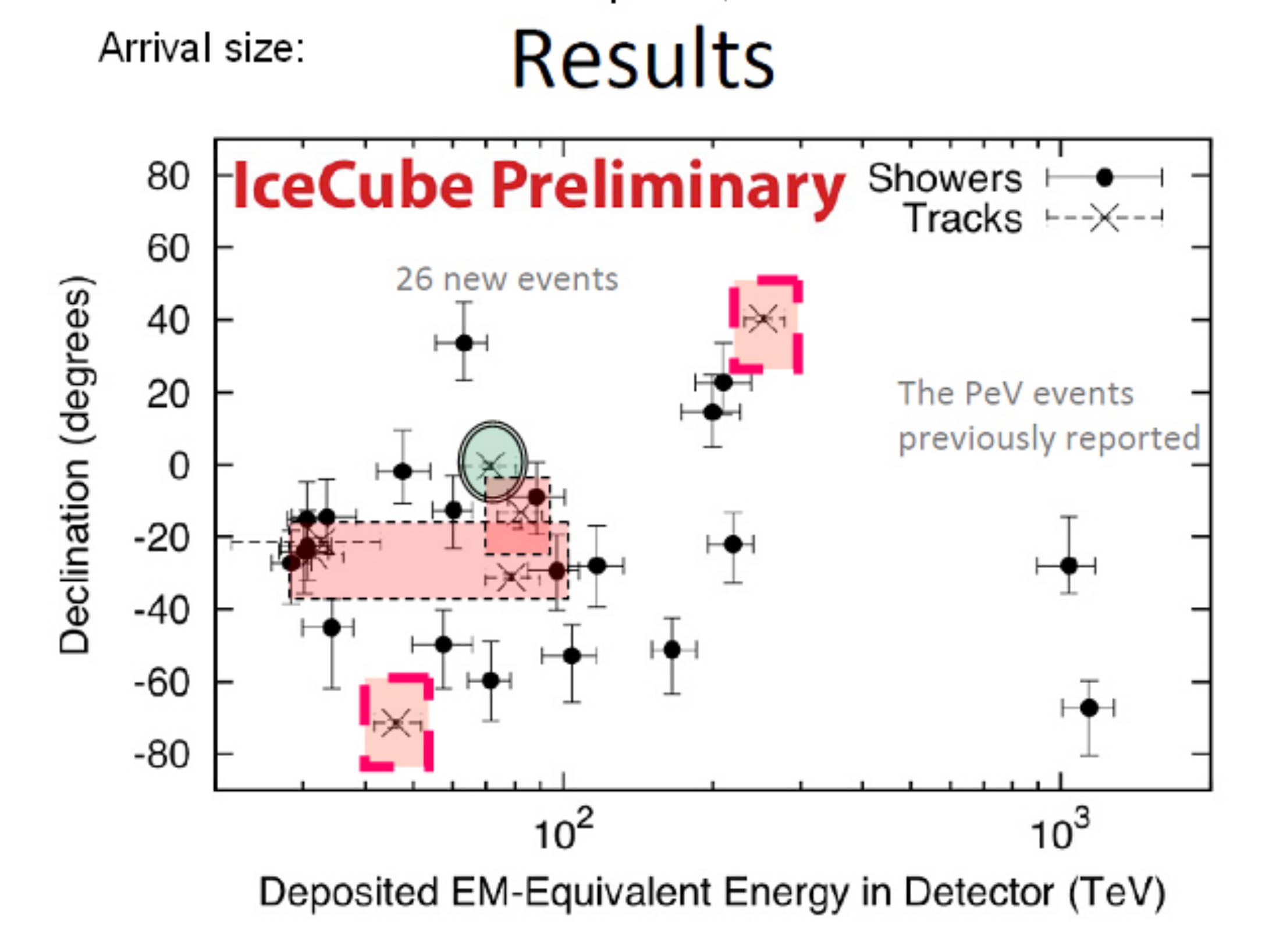}
\caption{Reinterpretation of each ICECUBE (see \cite{Klein-2013}) neutrino event nature: the hard dashed red boxes are too inclined to be atmospheric; a group of less inclined dashed downward events in square box might be also extraterrestrial; the very horizontal muon track in  a green circle is very probably the unique guaranteed atmospheric event, very probably ($\frac{7.4}{1}$) made by $K^{\mp}$ than by $\pi^{\mp}$ decay in flight} \label{Fig4}
\end{figure}

\begin{figure}[hbt]
\includegraphics[scale=.352]{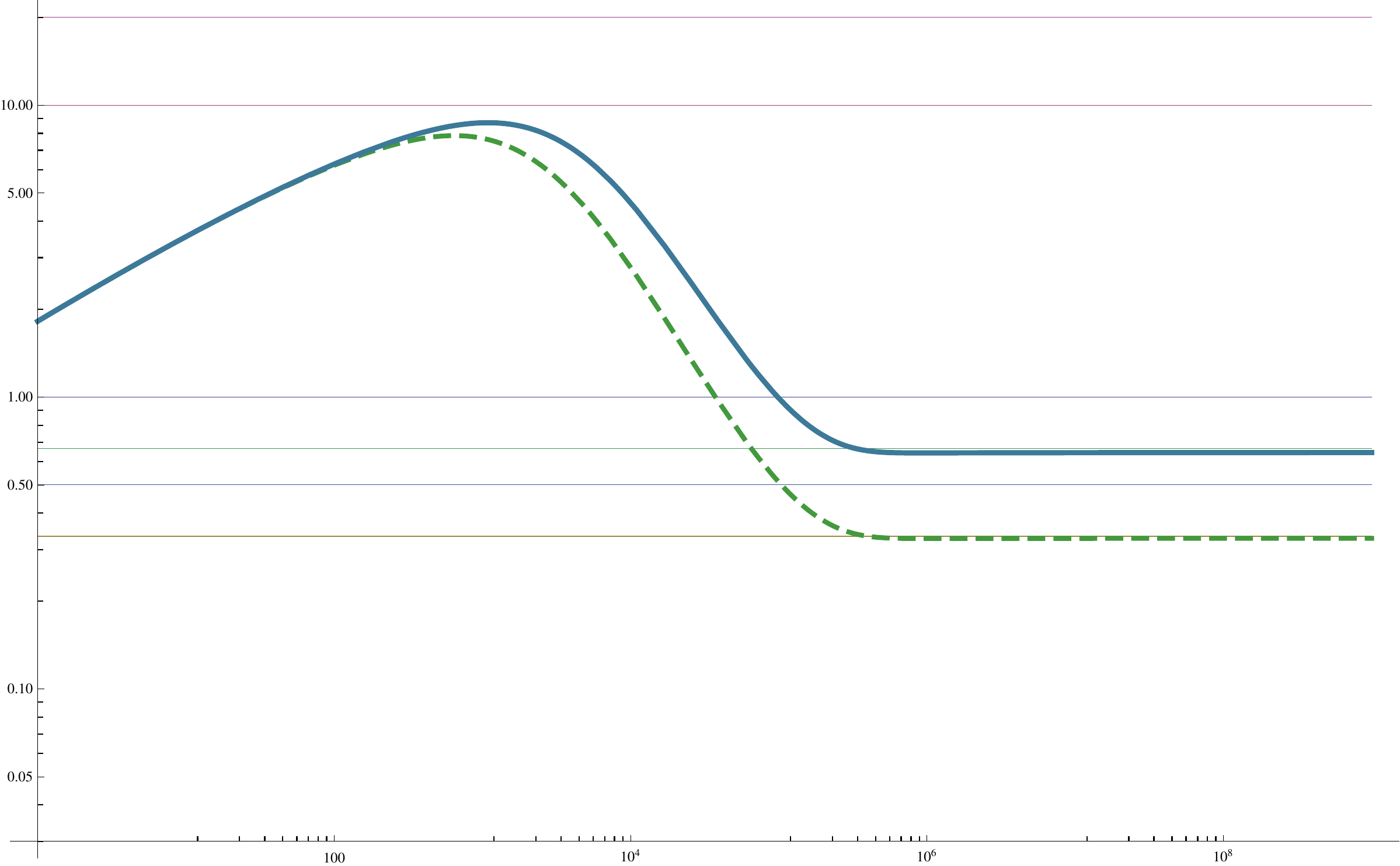}
\caption{The  needed forecast flavor ratio variability, to fit cascades and track ICECUBE events, a ratio between  the number of ($\nu_{\mu}$) tracks over cascade events (due to both flavor ($\nu_{e}$, $\nu_{\tau}$)) derived from ICECUBE data for low and highest energy events, for the most realistic fitting model considered in the text} \label{Fig3}
\end{figure}

   It should remind that at GeVs energies the $\mu$ pairs are widely deflected by geomagnetic fields; therefore there is also a local
   zenith and azimuth anisotropy somehow related to the historical geomagnetic (B.Rossi) cut-off in cosmic rays.  Moreover the
   atmosphere width  depend on the zenith angle geometry as well as on  the muon-pion-Kaon decay distances: therefore at TeVs energies the
    zenith anisotropy is remarkable. We shall concentrate to the average flux value.

Moreover at hundreds GeV muon neutrinos cannot oscillate much into tau states because of longer and longer oscillation distances (respect to Earth size). Therefore muon signals above tens GeV dominates again over electron ones while electron neutrino suffer as mentioned of the difficulty of energetic relativistic muons to decay. The pions and Kaons have a lifetime two order of magnitude shorter than muons and the Kaon mass makes the difference; therefore at two order of energy higher, at TeVs, also $\nu_{\mu}$ begin  to suffer of a pion-Kaon survival by a linear (with energy) suppression, and both $\nu_{\mu}$, $\nu_{e}$ flavor follow a harder spectra than parental cosmic rays with an experimental exponential index  $3.7$ as one would suggest, but at different ratio among the two flavors with an enhanced for  $\nu_{\mu}$ over $\nu_{e}$  by nearly an order of magnitude.
The atmospheric neutrino process based on cosmic rays has been  understood and tested from GeVs up to few tens TeV energy  also by recent ICECUBE results, see Fig.\ref{Fig4}. However, as we shall suggest, there is a possible overestimation of atmospheric neutrino flux at these highest energies. The general average behavior of the muon (added to its anti-muon) $\nu_{\mu}$ energy flux $ \Phi_{\nu_{\mu}}$
  between GeV up to PeV in average, ignoring here the muon-tau flavor detailed oscillations \cite{Fargion-2012}, may be approximated as follow to trace ICECUBE data :

                   $$  \Phi_{\nu_{\mu}} =  [(\Phi_{1_{\nu_{\mu}}})^{-1} + (\Phi_{2_{\nu_{\mu}}})^{-1}]^{-1} + \Phi_{0_{\nu_{\mu}}}$$
                   $$\Phi_{\nu_{\mu}}\equiv E_{(\nu_{\mu})}^{2}\frac{dN_{\mu}}{dE_{\mu}}$$
                   $$\Phi_{1_{\nu_{\mu}}} =5.0 \cdot 10 ^{7}\cdot (\frac{E_{\nu}}{GeV})^{-1} eV cm^{-2}\cdot s^{-1}\cdot sr^{-1}$$
                   $$\Phi_{2_{\nu_{\mu}}} =9 \cdot 10 ^{4}\cdot (\frac{E_{\nu}}{TeV})^{-1.7} eV cm^{-2}\cdot s^{-1}\cdot sr^{-1}$$
                   $$\Phi_{0_{\nu_{\mu}}} =12\cdot e^{\frac{-E_{\nu}}{2 PeV}} eV cm^{-2}\cdot s^{-1}\cdot sr^{-1}$$

     These functions keep care of the observed atmospheric $\nu_{\mu}$, $\bar{\nu}_{\mu}$ behavior in GeV- PeV range, trying
     to fit the 28 events  hardening at hundreds TeV and the needed cut off of events at PeV (because of the absence of additional events with higher energy in PeV region
     events and  in particular of $\bar{\nu}_{e} + e \rightarrow W$, the enhanced
     Glashow resonance at $E_{\bar{\nu_{e}}} = \frac{M_{W}^{2}}{2m_{e}}\simeq 6.3 PeV$).

   As mentioned $\nu_{e}$ flux as being secondary of muons  becomes suppressed at higher energy (GeVs-TeVs), i.e.  the initial flavor ratio $  R_{\frac{\nu_{\mu}}{\nu_{e}}} = \frac{\sum N_{\mu}} {\sum N_{e}} \simeq 2$
   for atmospheric  terrestrial neutrinos  becomes larger leading to a factor
       $$ R_{\frac{\nu_{\mu}}{\nu_{e}}} = \frac{\sum N_{\mu}} {\sum N_{e}} \simeq 20$$ at TeV energies as shown in Fig.\ref{Fig6}.

\begin{figure}[hbt]
\includegraphics[scale=.44]{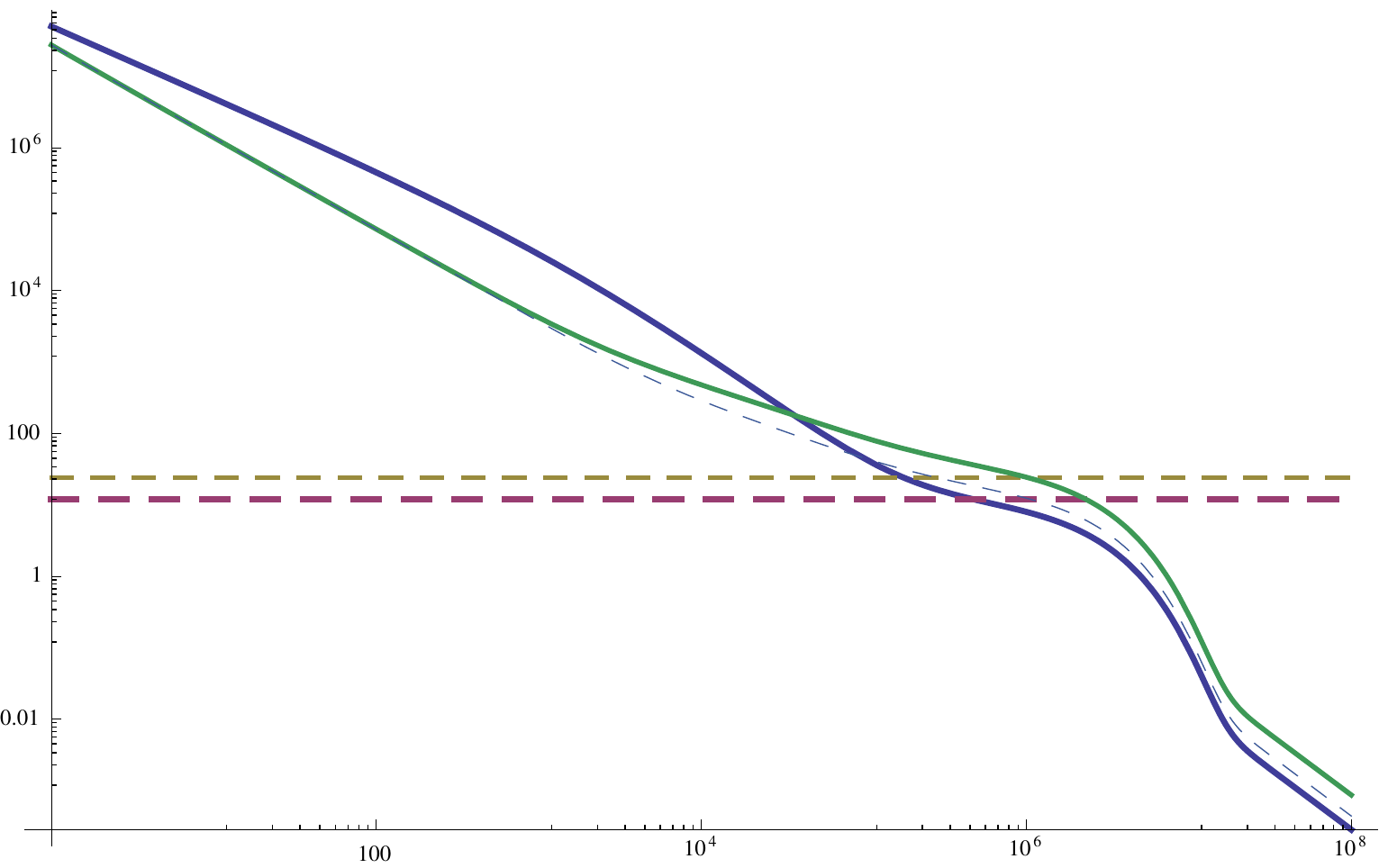}
\caption{Energy Fluency for $\nu_{e}$,$\bar{\nu}_{e}$,$\nu_{\tau}$ $\bar{\nu}_{\tau}$ flavor $\Phi_{\nu_e}$,$\Phi_{\nu_{\mu}}$ in $eV cm^{-2} s^{-1}sr^{-1}$ unity, as a function of the neutrino energy in GeV within a log-log graph. Note that the horizontal twin dashed lines stand for the observed fluency at highest  ICECUBE energy for one or two flavor. The thin dashed curve describe the role of one (of the two) showering flavor fluency by present description model } \label{Fig5}
\end{figure}

\begin{figure}[hbt]
\includegraphics[scale=.31]{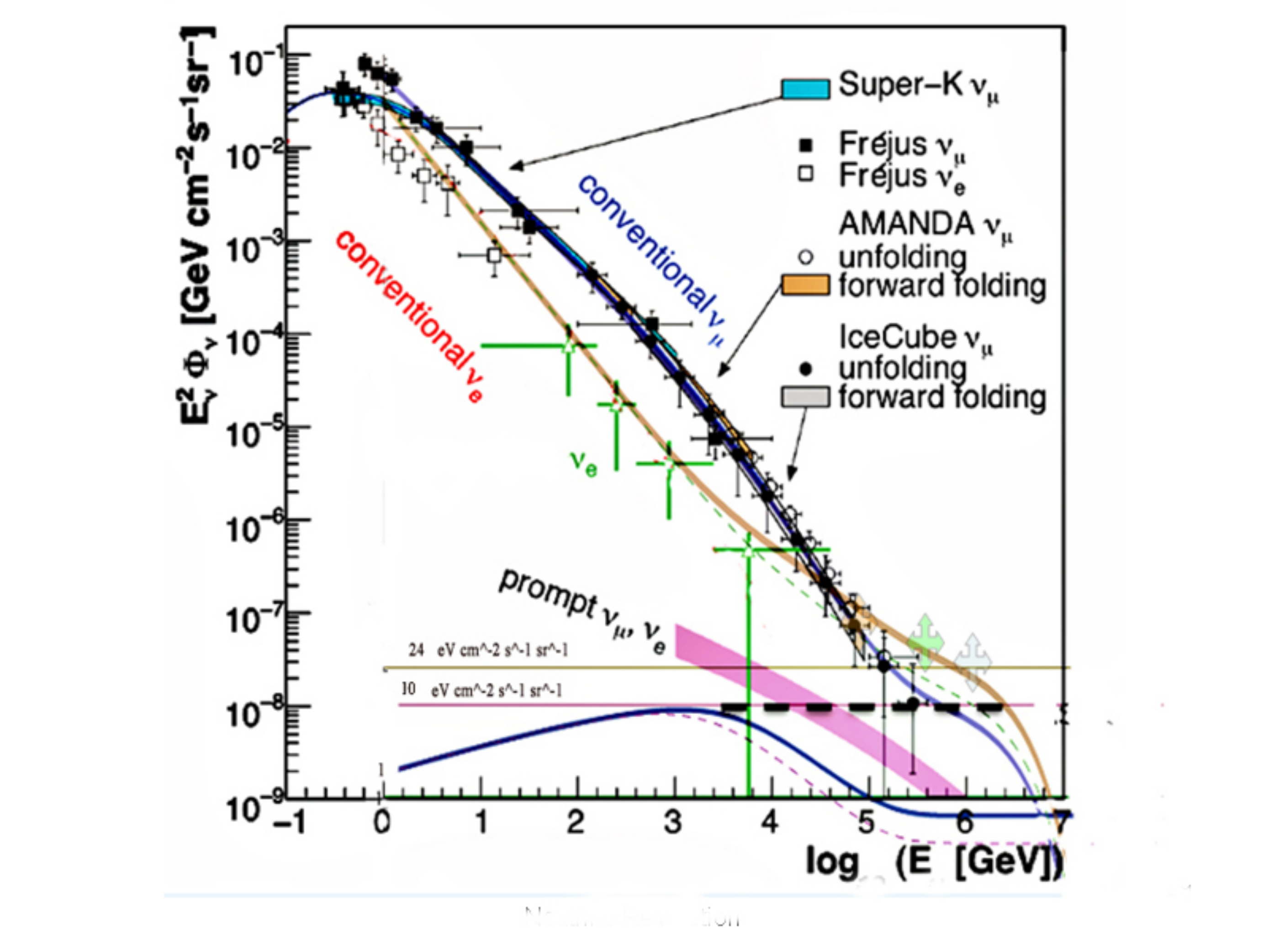}
\caption{Energy fluency $\frac{dN_{\nu}}{dtdAd\Omega}\cdot{E_{\nu}^{2}}$ for ($\nu_{e}$, $\nu_{\tau}$) showering cascades (green crosses) respect to same fluency for ($\nu_{\mu}$) tracks (blue continues curve) for old SK,Frejus and AMANDA as well as the recent  ICECUBE (green crosses by Deep Core detector) results, as a function of the energy in logarithmic-logarithmic scale \cite{Klein-2013}. The last 28 highest energy event are somehow displayed . All data fit \cite{Klein-2013}, \cite{Gaisser02}, \cite{Enberg08} followed by our curve model to explain the ICECUBE neutrino events. In the same figure at low figure area it is shown the flavor ratio as a function of the energy; near hundred TeV \emph{the sorpasso} takes place} \label{Fig6}
\end{figure}

\begin{figure}[hbt]
\includegraphics[scale=.50]{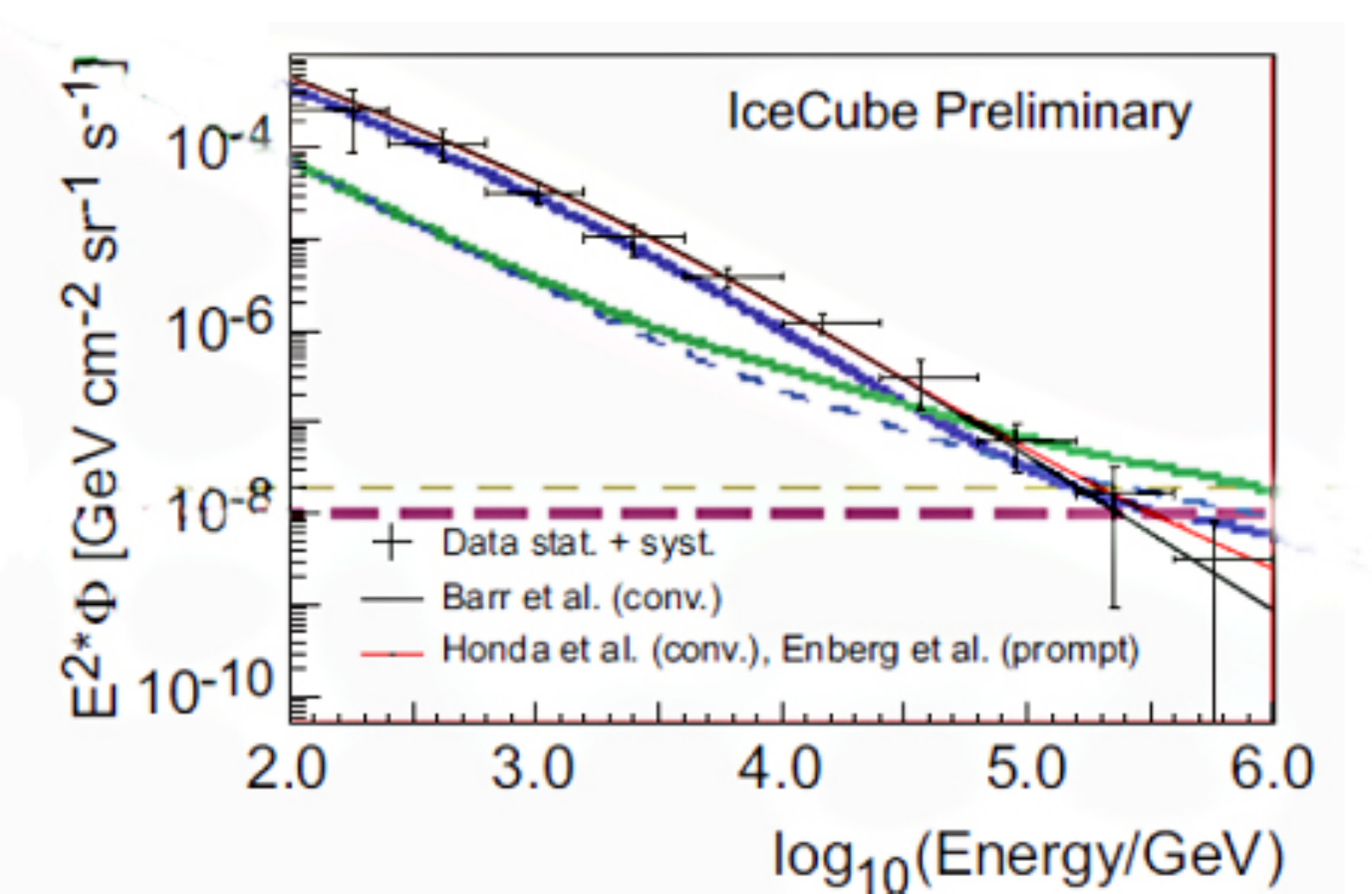}
\caption{Particular ICECUBE $\nu_{\mu}$ $\bar{\nu}_{\mu}$ flux events   \cite{Klein-2013} overlap by our curve fit; here we do not show the highest energy events shown in previous figure whose asymptotic fluency follow a dashed or bold dashed lines} \label{Fig7}
\end{figure}

        This occurs because we  begin from a flavor $\nu_{e}$  $\nu_{\mu}$ $\nu_{\tau}$ equal $1:2:0$ and we end (see Fig. \ref{Fig6}) at TeVs to
  a flavor ratio  $\nu_{e}$  $\nu_{\mu}$ $\nu_{\tau}$ equal $\simeq\frac{1}{20}:1:0$ (we neglect here the rare $\nu_{\tau}$ appearance by mixing at those energies).  Therefore muon atmospheric neutrinos are ruling over electron (cascade showers) at high (TeVs) energies.
 At higher energies, where the 28 events occur, (hundred TeV), Pion and Kaon decays (as well as muon late decay) are all too boosted and long life to take place; they mostly interact with nucleons  leading to an additional suppression in the spectra for both muons and electron $\nu_{e}$,$\bar{\nu}_{e}$,$\nu_{\mu}$,$\bar{\nu}_{\mu}$; at horizons their signal is crossing longest distances and they may survive  enhanced by nearly an order of magnitude respect vertical events. At highest (hundred TeV or few PeV) energies the rare very unstable charmed mesons  may decay faster, with little absorption in atmosphere, decaying equally into  $\nu_{\mu}$$\bar{\nu}_{\mu}$ and  $\nu_{e}$,
      $\bar{\nu}_{e}$ flavors, shining  nearly isotropically (out, of the upward Earth opacity): they may be the best source of  terrestrial highest energy neutrino; however these signals may provide only a small fraction ($1.5$)  among the $28$ ICECUBE events \cite{Klein-2013}.
       To describe in the following these observed $\nu_{\mu}$$\bar{\nu}_{\mu}$ and  $\nu_{e}$,$\bar{\nu}_{e}$ flux  we assumed an averaged $ \Phi_{\nu_{\mu}}$  shown  in Fig.\ref{Fig4}, Fig.\ref{Fig5}, Fig.\ref{Fig6}, Fig.\ref{Fig7}.

The muon and electron flavor ratio $R_{\frac{\nu_{\mu}}{\nu_{e}}}$ metamorphosis as a function of the neutrino energy $E_{\nu}$ in GeV unity, is shown in Fig \ref{Fig3}
 approximated by  a simple analytical law that may meet at best the flavor  behavior and ICECUBE records. The main  functions structure are three: $R_{1}$ function that describes the rise of the extraterrestrial neutrino component, whose cascades (by one $\nu_{e}$ or by two $\nu_{e} + \nu_{\tau}$) fit the ICECUBE data, $R_{2}$ keep memory of largest flavor difference, $R_{3}$ describes the fast extraterrestrial appearance:
          $R_{1}= 2 \cdot E^{\frac{1}{3}} $, $R_{2}= A  $;
          $R_{3}= \frac{3}{2} \cdot(1- e^{-\frac{E_{\nu_{mu}}}{10^{5} GeV}}) $

   $$ R_{\frac{\nu_{\mu}}{\nu_{e}}} =[R_{1}^{-1}+ R_{2}^{-1} + R_{3}]^{-1}  $$
   $$ R_{\frac{\nu_{\mu}}{\nu_{e}}} =[R_{1}^{-1}+ R_{2}^{-1} + 2 \cdot R_{3}]^{-1}$$

   Where   the constant $A= 20$, the neutrino energy $E_{\nu_{mu}}$ is in GeV unity.
    The last two laws are for one showering flavor or for two,
     namely the unique $\nu_{e}$ or both $\nu_{e}$ and $\nu_{\tau}$, as long as
     the $\nu_{\tau}$ energy is below the few PeV where double bang signature
     might be noticed.
     We wish to remind that up to now there are not yet any
     evidence of such twin event due to $\nu_{\tau}$ first nuclear interaction and later  on ${\tau}$ decay. Surprisingly, possibly, also for the third announced PeV event.
      Anyway the exponential decay term in $R_{3}$ stand for a rapid change of flavor at hundred TeV. Finally
    $$     \Phi_{\nu_{e}}\simeq   \Phi_{\nu_{\mu}}\cdot [R_{\frac{\nu_{\mu}}{\nu_{e}}}]^{-1}  $$

  This flavor ratio evolution with energy $R_{\frac{\nu_{\mu}}{\nu_{e}}}$ is shown in (see Fig.\ref{Fig3} for one (dashed) or two flavors) where both muon and electron atmospheric neutrino evolution is derived from the data (see Fig.\ref{Fig4}) combined in the model above for the fluency in the following figure (see Fig.\ref{Fig5}). The
  model and the data are combined in (see Fig.\ref{Fig6}),and in more detail are shown in last Fig.\ref{Fig7}. The dashed curve stand for one electron shower, the combined electron and tau are described by the continuous line.

\section{An atmospheric prompt neutrino  role?}
The  energetic prompt atmospheric neutrinos cannot oscillate much inside the terrestrial size and therefore they may keep the same primary ratio $\frac{N_{\mu}}{N_{e}}= 1$; $\nu_{\tau}$ flavor flux is an order of magnitude smaller and maybe neglected. The expected prompt neutrino flavor ratio is $\nu_{\tau}:\nu_{\mu}:\nu_{e}\simeq 0:1:1$. This ratio is much more suitable in explaining the sudden muon reduction than conventional ones. Therefore one may wonder if the present flavor paradox maybe solved at best within a total prompt  atmospheric neutrino scenario; the probability that  among $28$ events only $7$ are muon track is $P_{Prompt}$ $ = [{\frac{n!}{(n-k)!k!}\cdot p^{n-k}\cdot q^{k}}]$,  where $ p=q= \frac{1}{2}$ and $n= 28$, $k=7$, or just $P_{Prompt} = 0.44\%$, while the cumulative probability that among $28$ events only $7$ or less are muon tracks is just $P_{Prompt} \leq 1.3\%$. Therefore a global prompt atmospheric neutrino role for ICECUBE events cannot solve much the flavor paradox.

\section{Hiding atmospheric role within a 30-60 TeV  band?}
 One may imagine, following a referee suggestion,   that most
   atmospheric neutrino are hidden in the lower range,
       within 30-60 TeV energy band; among  the higher energy 17 events (4 tracks, 13 showers)
   above 60 TeV   only 4 events are muons tracks, implying,
    a negligible or none atmospheric pollution.

   However a more carefull analysis of each event seem still to disfavor
    this new interpretation even
   above 60 TeV: among the 17 events two (event ID 5, ID 5 23) \cite{Science-2013}, seem very probable atmospheric ones
   because low energetic and very horizontal, leaving just 13 showers versus 2 tracks
    to be explained.
The probability for such  situation to occur is below $22\%$ (assuming  a probability of $\frac{1}{4}$ for track/shower ratio )
    or below $7.7\%$ (assuming  a more realistic  probability of $P= \frac{1}{3}$ for track/shower ratio due to the negligible role of NC)\cite{Vissani-2013}.

   Indeed the event n. 5 \cite{Science-2013}, a clear muon neutrino track, is nearly horizontal (-0.4 degree)
   and low energetic therefore it is very probably of atmospheric nature; also events 23 (-13.2 degree) is  like that.
   But, once again,  the puzzle survive even more remarkable
   for those events at lower energies (60 TeVs-30TeV) hidden into a limbo: (11) events
   that (in principle) should be  mostly polluted by
   atmospheric  muons and their atmospheric neutrinos ($10.6^{+5.0} _{3.6}$).
   In this view this region (30TeVs-60 TeV) one observe  an unexpected (8) showers and 3 tracks
    versus a characteristic  (in early TeV-10 TeV range by Deep Core) negligible showering rate
   (1 over 3 assuming NC );   for well known reasons it is
   better to say that the expected shower versus tracks in atmospheric dominated range is nearly 1 over 10 \cite{Vissani-2013}).
   The discontinuity in the flavor metamorphosis inside  such a narrow energy range is even more surprising
   in  view of the fast flavor variability (from $TeV-30 TeV$) while showing in the same range a steady muon neutrino flux decay.
  Therefore if the paradox is reduced in the high energy range  it is strongly enhanced in lowest ones.

  In conclusion to solve the puzzle there may be a growing
   role of extra-terrestrial neutrino all along the 30-TeV-PeV energy, leaving a marginal atmospheric $\nu$ role; such transition
   should be ruling more and more (1 over ten) on atmospheric ones, suggesting that the dominance occurs
   already at earlier energies,  around 10  TeV, as suggested in our articles;
   Otherwise, as the new proposal and the connected question marked in the title, a sudden flavor change at 30-60 TeV
   seem unexplained: P(k=8,n=11) $= 7.45\cdot 10^{-3}$ assuming  a ratio $1/3$ for NC over a complete CC rate for most  neutrino;
  assuming a more realistically probability $P= \frac{1}{10}$ for NC (at given energy) over $P= \frac{9}{10}$ for CC interactions
    by atmospheric muon neutrino \cite{Vissani-2013}, the probability to occur is almost vanishing:
    $P(k=8,n=11) =$ $1.2 * 10^{-6}$: there is no  room for an atmospheric hidden role in
   a $30-60$ TeV energy band. Otherwise the puzzle is even greater.

\section{Conclusions: Extraterrestrial neutrino at tens TeV}
Extraterrestrial neutrinos  (by AGN,GRBs by jets or by UHECR via GZK cut off or their decay in flight, by prompt charmed interaction and decoherence) may also rise at highest energy. Their  rate, almost comparable in each flavor after oscillation and mixing may pollute or even rule the neutrino highest energy edges.
Let us remind that the de-coherence of the $\nu$ flavors after oscillation depends on the primary rate and the flavor matrix discussed in recent articles \cite{Fargion-2011}, \cite{Fargion-2012}
$ P_{\alpha \beta} = \sum_{i=1}^3 U_{\alpha i}^2 U_{i \beta}^2  $.
For instance, for  the GZK neutrinos, due to UHECR  scattering on relic photons, their flavors $\nu_{\tau}:\nu_{\mu}:\nu_{e}$ are  born by photo-pion decays usually  as $0:2:1$ ; because oscillation they reach us as $\simeq 1:1:1$. The consequent probability to occur as observed by ICECUBE by $21$ cascades and $7$ tracks is $P_{1:1:1}= 10.8\%$, a realistic range.  Moreover the astrophysical neutrinos may also rise by proton proton scattering leading once again to a comparable final flavor population; also astrophysical UHE neutrino maybe also be born as a prompt ones within  $\nu_{\tau}:\nu_{\mu}:\nu_{e}\mapsto$  $0:1:1$ reaching us after decoherence,  at a ratio $\nu_{\tau}:\nu_{\mu}:\nu_{e}\mapsto 0.283:0.311:0.406$; therefore   $\nu_{e}$, $\nu_{\tau}$ showering will already rule (ignoring the neutral current role) at $68.9 \%$ ratio over all the events (the muon tracks take place within a probability of $31.1 \%$), more consistently with the observed ICECUBE showering-tracks $3:1$ ratio. If one takes into account also the  neutral current (leading to additional one third cascades at lower energies), the final ratio become: $\frac{P_{track}}{P_{Cascade}}\mapsto \frac{0.763}{0.238}= 3.28$, a value even more consistent with data (if at least one or two muons are indeed atmospheric ones).
  The probability that ICECUBE events are born  within such a prompt extraterrestrial origination is even more favorite because  $P_{0:1:1}\mapsto 0.406:0.311:0.283$ while previous $\simeq 1:1:1$\emph{ democratic }flavor distribution: $P_{1:1:1}= 10.8\%$,  $P_{Prompt} = 13.34\%$. Finally  we considered also a beta decay role where UHECR are radioactive heavy or lightest nuclei whose secondaries  neutrino feed also astrophysical neutrino flux. The flavor ratio in such a beta decay flux scenario, is originally $\nu_{\tau}:\nu_{\mu}:\nu_{e}\simeq 0:0:1$ but the final  flavor abundance after de-coherence  becomes $\nu_{e}$ dominated $\nu_{\tau}:\nu_{\mu}:\nu_{e}\simeq 0.1883:0.2643:0.5473$. In this case the final rate probability for an  electron or a tau whole showering  is as large as  $73.5\%$ for cascade over a much rarer $26.4\%$ muon track signature; if one takes into account also NC cascades this ratio become  slightly larger, $\frac{P_{Cascade}}{P_{track}}\mapsto \frac{0.799}{0.198}= 4.03$ a ratio at best compatible with the observed ones in ICECUBE assuming just $5$ extraterrestrial event are muons: $\frac{N_{shower}}{N_{all}}\simeq \frac{4}{5}$.

\emph{In conclusion} contrary to present ICECUBE understanding the recent $28$ highest energy neutrinos \cite{Klein-2013} might be  \emph{mostly all of extraterrestrial origin}; atmospheric ones must have, we believe, a softer spectra than expected \cite{Klein-2013}, even a factor ten. A few (one-two) may be real atmospheric $\nu_{\mu}$ one as the horizontal ones (see Fig.\ref{Fig6}, Fig. \ref{Fig4}); the other  $\nu_{\mu}$, $\nu_{\tau}$, $\nu_{e}$ should be already extraterrestrial signals showering cascades more than tracing muon tracks. Their flavor \emph{revolution} require a transition from atmospheric to extraterrestrial ones already about ten TeV. This imply a relevance in clustering and correlation map at a few ten TeV energies. The muons crossing flux by $\nu_{\mu}$ at horizons or at vertical (where Earth opacity is still negligible) may be a key meter to verify the expected overabundance of horizontal atmospheric neutrinos  versus vertical ones (at least an order of magnitude) within  $ 5^{o}$ of the upward horizons respect vertical ones , contrary to our foreseen mostly isotropic extraterrestrial flux. The number of such crossing events should be about few or several tens a year. The tens TeV neutrino sky and clustering with the its flavor \emph{revolutions} is a crucial cornerstone to solve ICECUBE muon paucity puzzle. With more events a sharper sky  map  may better address us to  the eventual galactic or  extragalactic origination. Antares, as large as twice Deep Core, might contain also rare cascades at few and tens TeVs  testing somehow the early flavor metamorphosis.
\subsection{ Dedicated to 16 October 1943} This article is devoted to the memory of all the Jew deported to  Auschwitz 70 years ago, on 16 October 1943, from Rome.

\clearpage

\end{document}